Article

# *Acinetobacter baumannii* Deactivation by Means of DBD-Based Helium Plasma Jet

**Panagiotis Svarnas** [1,*], **Anastasia Spiliopoulou** [2], **Petros G. Koutsoukos** [3,4], **Kristaq Gazeli** [1,†] 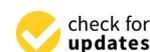 **and Evangelos D. Anastassiou** [2]

1. High Voltage Laboratory (Plasma Technology Room), Department of Electrical and Computer Engineering, University of Patras, 26 504 Rion, Greece; kristaq.gazeli@u-psud.fr
2. Department of Microbiology, School of Medicine, University of Patras, 26 504 Patras, Greece; spil@upatras.gr (A.S.); anastas@med.upatras.gr (E.D.A.)
3. Department of Chemical Engineering, Section of Chemical Technology and Applied Physical Chemistry, University of Patras, 26 504 Rion, Greece; pgk@chemeng.upatras.gr
4. Institute of Chemical Engineering Sciences (ICES)-FORTH, 26 504 Rion, Greece
* Correspondence: svarnas@ece.upatras.gr
† Current Address: LPGP, CNRS, Univ. Paris-Sud, Université Paris-Saclay, 91405 Orsay, France.

**Abstract:** Acinetobacter baumannii is a typically short, almost round, rod-shaped (coccobacillus) Gram-negative bacterium. It can be an opportunistic pathogen in humans, affecting people with compromised immune systems, and it is becoming increasingly important as a hospital-associated (nosocomial) infection. It has also been isolated from environmental soil and water samples. In this work, unlike conventional medical methods like antibiotics, the influence of atmospheric-pressure cold plasma on this bacterium is evaluated by means of a colony count technique and scanning electron microscopy. The plasma used here refers to streamers axially propagating into a helium channel penetrating the atmospheric air. The plasma is probed with high resolution optical emission spectroscopy and copious reactive species are unveiled under low-temperature conditions. Based on the experimental results, post-treatment (delayed) biochemical effects on Acinetobacter baumannii and morphological modifications appear dominant, leading to complete deactivation of this bacterium.

## 1. Introduction

Atmospheric-pressure plasma jets (APPJs) are novel micro-discharges, which allow for the effective formation of user-friendly non-thermal plasmas under ambient-air conditions. They are mainly produced by means of Dielectric-Barrier Discharge (DBD) reactors of different electrode configurations and dielectric materials [1]. These devices are electrically driven mostly by sinusoidal and pulsed high voltage power supplies of audio frequencies [1,2]. The use of the dielectric barrier placed between the electrodes and the short duration of the applied voltage (in the case of pulsed power supply) limit the discharge current. The transition to arc is thus avoided [1]. In general, the plasma is formed in noble gases, like He or Ar [1,2], due to their higher ionization efficiency with respect to that of the ambient air under the same reduced electric field [2,3]. The discharge comprises a DBD region, a displacement along the dielectric tube and a penetration into the atmospheric air at distances up to a few centimeters [4–6].

Although APPJs appear as continuous bright "plumes" to the naked eye, in fact they consist of ultrafast discrete ionization fronts widely known as "plasma bullets" [3,7,8] and, more recently, as "guided streamers" [2–4,9]. The latter term mirrors better the physical nature of APPJs. The streamers propagate inside the thin noble gas channel (few millimeters or lower in diameter) with velocities in the order of tens of km s$^{-1}$ [1–4,10–12], which are two-three orders of magnitude higher than that of the noble gas itself [4]. This means that the plasma formation outside the reactor is an electrically-driven phenomenon, which is strongly coupled with the noble gas dynamics and its mixing with the ambient air [13–15].

During the last fifteen years, numerous works have been devoted to the development and systematic characterization of APPJs, not only for fundamental research purposes [2–5,16–18], but also for their implementation in novel technological domains (material-surface processing [19–22], food industry and agriculture [23], and biomedicine [24–31]). These applications are promoted due to APPJ special features, like relative low construction cost, portability, μm-to cm-scale application, low temperature of the gaseous medium (even near to the room one for well-defined operational windows [5,32]), and mostly due to their operation at atmospheric air which permits the effective production of various (re)active species.

These species refer to excited neutrals (atoms/molecules) including long-living metastables, ground state reactive radicals, electrons over a wide range of energy (from eV to tens of eV), atomic/molecular ions, photons (UVA, UVB, UVC, visible) etc., combined with electromagnetic and gas flow fields. Specifically, in the plasma biomedicine area, the synergetic action of the above plasma-induced components is responsible for the effects observed on discrete living samples. Especially, the so-called Reactive Oxygen and Nitrogen Species (RONS) [33] play a crucial role in wound healing and tissue regeneration, blood coagulation, sterilization, decontamination, plasma-mediated cancer therapy and killing of micro-organisms. Besides, metastable states [16,34] act as energy reservoirs by stocking energy for relatively long time, supporting thus excitation and ionization processes outside the reactor where biological specimens are being placed. APPJ potential reactivity is discussed below in more detail.

Among other topical applications of APPJs in biomedicine, particularly interesting (and increasingly studied) are their interactions with micro-organisms. Nosocomial or health-care associated infections, including bacteremia, represent a leading cause of death worldwide [35]. Most importantly, the irrational and excessive consumption of antimicrobial agents has led to multi- or even pan-drug resistant bacterial strains, leaving no available treating options in the armamentarium of clinicians. Looking to the twilight of the era of antibiotics, other treatment approaches appear thus as an appealing as well as an imperative alternative.

Towards this direction, the efficiency of APPJs to the inactivation of different kinds of bacteria and fungi has been tested widely. Interrogated microorganisms in different reports are given below, just to mention a few: Escherichia coli, Staphylococcus aureus and Pseudomonas aeruginosa [36], Escherichia coli, Bacillus subtilis and Pseudomonas aeruginosa [37], Escherichia coli, Bacillus atrophaeus and Staphylococcus aureus [38], Escherichia coli [39], "skin flora" (mix of bacteria) [40], Escherichia coli [41], Escherichia coli and the derived lipopolysaccharide [42], Enterococcus faecalis and Staphylococcus aureus [43], Escherichia coli and methicillin-resistant Staphylococcus aureus [44], Staphylococcus aureus [45], Candida biofilms [46], Bacillus subtilis spores [47], bacteria of skin samples consisting of Staphylococci, Streptococci and Candida species of yeast [48,49], Escherichia coli [50], methicillin-resistant Staphylococcus aureus, Pseudomonas aeruginosa and Candida albicans [51], biofilms produced by Escherichia coli, Staphylococcus epidermidis and methicillin-resistant Staphylococcus aureus [52], Escherichia coli, Staphylococcus aureus, Micrococcus luteus, Bacillus natto, Bacillus subtilis and Bacillus megaterium [53], Bacillus stearothermophilus spores [54], Bacillus globiggi [55], Bacillus subtilis and Eschelichia coli [56], Streptococcus mutans and Lactobacillus acidophilus [57], Escherichia coli and Enterococcus mundtii and Candida albicans [58], Escherichia coli and Bacillus subtilis [59], Geobacillus stearothermophilus [60], Bacillus atrophaeus, Geobacillus stearothermophilus, Salmonella enteritidis, Staphylococcus aureus, Candida albicans and Aspergillus niger [61], Escherichia coli [62], Escherichia coli and Staphylococcus [63], Staphylococcus aureus [64], Bacillus spores [65], Pseudomonas aeruginosa biofilms [66], Acinetobacter baumannii, Enterobacter cloacae and Staphylococcus aureus [67], Staphylococcus aureus, Pseudomonas aeruginosa, Acinetobacter baumannii, Candida kefyr [68], Escherichia coli, Staphylococcus aureus, Acinetobacter baumannii and Staphylococcus epidermidis [69], Acinetobacter baumannii (A. baumannii) [70] etc.

In particular, A. baumannii is an aerobic Gram-negative coccobacillus that causes nosocomial human infections, particularly in immune-compromised individuals. These infections can result in septicemia, meningitis, endocarditis, pneumonia, wound infection, and urinary tract infections. A. baumannii can colonize and survive for long periods of time on dry inanimate surfaces, such as hospital equipment. In addition to its intrinsic resistance to antibiotics and desiccation, A. baumannii is very prone to accumulating resistance mechanisms, especially in environments where antibiotic overconsumption is taking place [71]. Its interaction with APPs has attracted interest [67–70], since APPs could assist with the elimination of nosocomial infections associated with this bacterium, but still not many works on this topic can be found in the literature. Parkey et al. [67] used a single outer electrode plasma-jet reactor, Kolb et al. [68] a DC-biased micro-hollow cathode geometry, Ercan et al. [69] a floating-electrode DBD, and Ruan et al. [70] a multi-needle DC-driven plasma source. Thus, the reports on A. baumannii deactivation by APPJs are few and far between.

Unlike to the abovementioned excellent works, herein a coaxial DBD-based helium plasma micro-jet is evaluated against A. baumannii deactivation. This setup has been previously used for liposome [29,30] and human skin [31] treatment, under various conditions. Thus, here colony counting and scanning electron microscopy (SEM) provide insights into the plasma-induced damage of A. baumannii and it is shown that post-treatment (delayed) effects are dominant. Bacterial population is extensively diminished (reduction of seven orders of magnitude, without the need for co-injected air as in reference [67]) and remaining bacteria resemble ghost bacteria consisting of ruptured and destroyed cell membrane. Possible decontamination mechanisms, based on optical emission spectroscopy (OES) results of the present study and bibliography, are discussed.

## 2. Experimental Setup and Specimens

### 2.1. Plasma Setup

The specifications of the plasma-jet reactor employed here for the bacterium treatment and the involved physics for the same [29–31] or similar [4,5,12,32,72,73] coaxial DBD-based systems have been discussed in these previous reports of our group. The present operational window (see below) is selected as an effective one, based on those previous studies. Additionally, although we are aware that any aqueous solution may alter the plasma-jet features, acting practically as a third (liquid) electrode, we provide here conventional results from the free running plasma-jet case as indicative ones, whereas we focus our attention on the possibility of A. baumannii deactivation. The role of the bacterial suspension in the plasma features is a part of our ongoing research.

Briefly, as in Figure 1, a thin tungsten wire (diam. 0.125 mm) is inserted in an alumina capillary tube (inner diam. 1.14 mm). A brass hollow cylinder (length of 10 mm) is attached on the outer surface of the alumina in such a way to be extended up to the tip of the wire. The distance between the tip of the wire and the tube orifice is 20 mm. Helium gas (N50) is introduced in the tube with a controllable flow rate. The wire is biased by a laboratory-built high voltage power supply (10 kHz sinusoidal; 0–12 kV peak-to-peak) and the cylinder is grounded directly. By this configuration, radial DBDs are developed and axial "guided streamers" propagate [12,73] in the atmospheric air where helium diffuses. Here, the voltage and the flow are fixed at 11.5 kV and 2 slm, respectively, and a plasma "plume" of about 35 mm is achieved. Figure 2 presents a typical pair of the driving voltage and the generated DBD current waveforms, as they are recorded by means of a wideband high voltage divider (Tektronix P6015; DC–75 MHz, Beaverton, OR, USA) and wideband current monitor (Pearson Electronics 6585; 40 Hz–200 MHz, Palo Alto, CA, USA), respectively, on a digital oscilloscope (LeCroy

WaveRunner 44Xi-A; 400 MHz; 5 GSamples s$^{-1}$, Chestnut Ridge, NY, USA). Emissive species and gas temperature are identified by UV-visible, high resolution OES, as described in references [4,29–32]. In particular, the gas temperature is evaluated by comparing experimentally-determined rotational distributions of probe molecules with the corresponding theoretical ones calculated with a software developed previously [74]. The probed zone is 20 mm along the plasma-jet axis with respect to the tube orifice.

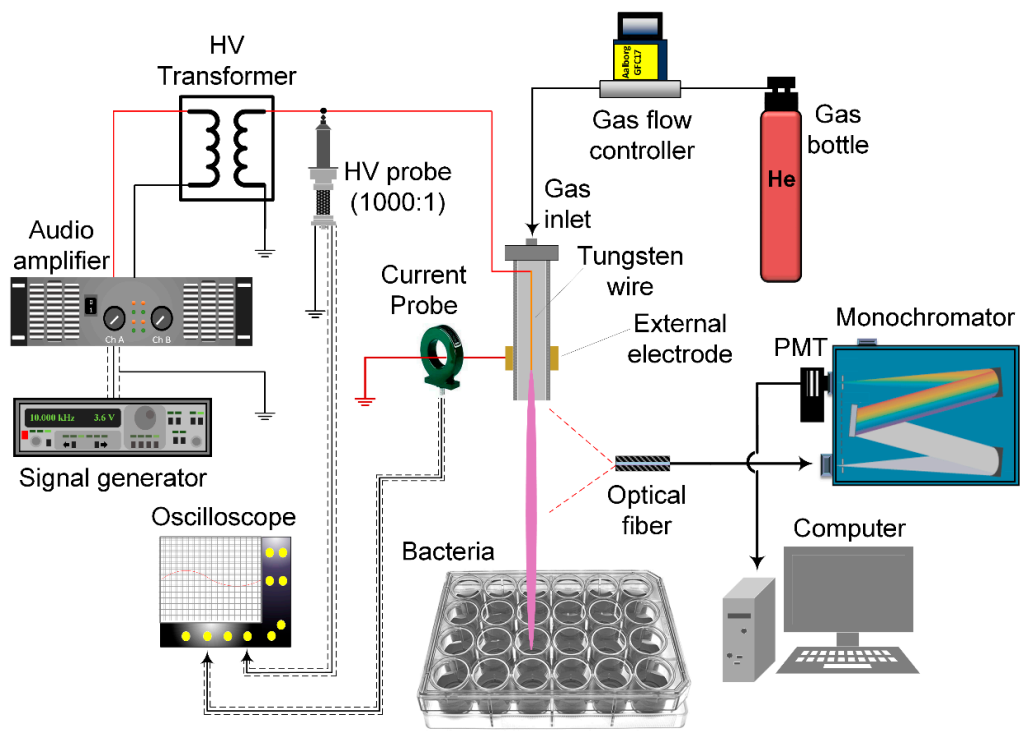

**Figure 1.** Conceptual view of the DBD-based plasma-jet experimental setup used here.

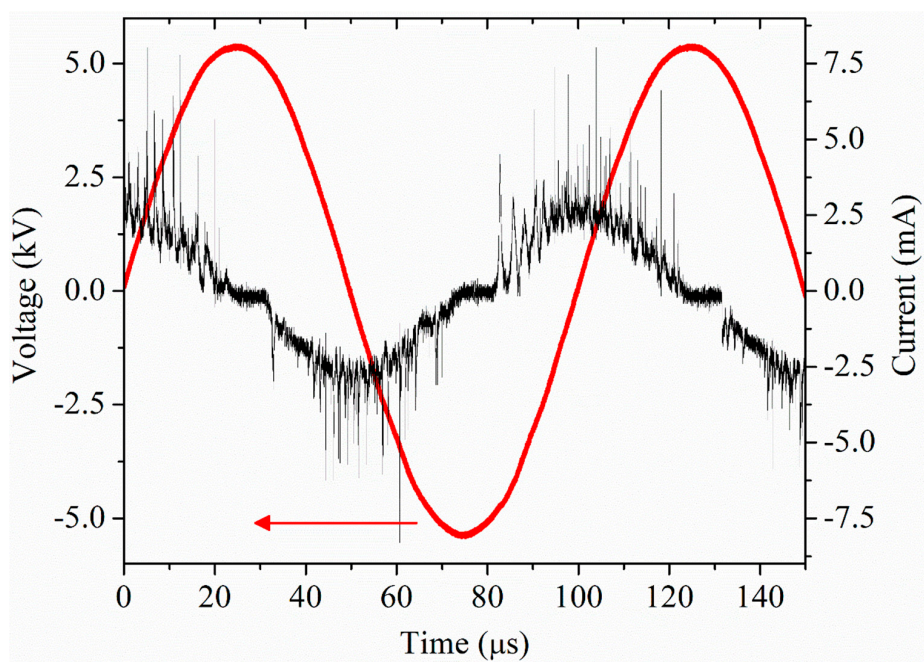

**Figure 2.** Representative driving voltage and DBD current oscillograms over one and half periods.

*2.2. Bacteria Preparation and Treatment*

The bacterium under study is *A. baumannii*. All strains are isolated in the Department of Clinical Microbiology, University Hospital of Patras, Greece, from the blood of patients treated for bloodstream infections and represent multi-resistant strains. All microbial strains used in this study are stored in TSB (Tryptic Soy Broth; BBL, Microbiology, Cockeysville, MD, USA) supplemented with 20% glycerol at −70 °C and, when needed, they are sub-cultured on blood agar plates (BioMérieux S.A., Marcy-l'Étoile, France) for 24 h at 37 °C.

Each bacterial isolate is suspended in PBS (Phosphate-Buffered Saline; pH 7.2) and photo-metrically adjusted to the desired concentration of $3 \times 10^6$ CFUs mL$^{-1}$ (Colony Forming Units mL$^{-1}$). 1 mL from the suspension is added in each of nine (three for each treatment time) wells of a 24-well flat bottom tissue culture plate (Sarstedt, Newton, USA). Bacterial suspensions are treated at 4 mm distance between the tube orifice and the surface of the aqueous solution for 0 (control samples), 10 and 20 min. The content of each of the nine wells is kept at 4 °C in separate sterile Eppendorf$^{TM}$ tubes.

Different dilutions of each suspension are sub-cultured at 2, 6, 24, 48 and 72 h upon plasma treatment on blood agar plates and incubated for 24 h at 37 °C. Corresponding bacterial concentration is thus estimated. The effect of the plasma treatment is evaluated depending on the number of live bacteria capable of forming colonies on solid media. Bacterial concentration at the content of the three wells that are not treated with plasma (0 min) is used as control reference. The effect of helium gas itself (plasma off) on the bacterial viability is evaluated by subjecting bacterial suspensions to 2 slm He flow. Negligible effect is measured with respect to control samples.

*2.3. Growth Inhibition Zone Experiments*

For growth inhibition zone determination, an overnight culture of bacteria is diluted in PBS at a concentration of $3 \times 10^7$ CFUs mL$^{-1}$ and 10 µL of the diluted bacterial suspension is then spread evenly on the surface of 7 mm MHA (Mueller Hinton Agar) plate. The plates are exposed (in duplicate) to the plasma for 30, 60, 120 and 240 s at distances of 10 and 15 mm from the reactor orifice. After exposure to the plasma, the plates are incubated at 37 °C for 24 h in a static incubator and the diameter of the bacterial growth inhibition zone on the agar surface is determined [75]. Since the inhibition zone circumference is irregular, the inhibition zone diameter is approximated by applying simple image processing techniques, i.e., zone edge detection with a Sobel-Feldman operator and optimum circular fit, using Matlab$^{TM}$.

*2.4. Scanning Electron Microscopy of Bacteria*

The specimens are prepared for morphological examination by Scanning Electron Microscopy (SEM). The microscope is a LEO-ZEISS Supra 35VP instrument equipped with an Energy-Dispersive X-ray (EDX) microanalysis unit (Bruker), allowing for elemental semiquantitative analysis of the specimens. The content of each well is filtered on 0.45 µm filters (Sartorius, Göttingen, Germany). Withheld bacteria on the filter surface are fixed in a solution of 2.5% glutaraldehyde overnight at 4 °C, carefully washed two times for 10 min in PBS and dehydrated sequentially in 10%, 30%, 50%, 70%, 80% and 90% ethanol solutions for 10 min each and eventually in ethanol 100% for $2 \times 10$ min, as previously described [46,76] but with small alterations. Eventually, a thin film of about 4 nm of gold is coated on the filter-bacteria by means of a BAL-TEC SCD-005 sputter coater.

### 3. Results and Discussion

Figure 3 presents the number of survived bacteria in CFUs as a function of the time elapsed upon the termination of 10 and 20 min treatment (i.e., post-treatment). The evolution of the untreated sample (control) is shown as well for comparison reasons. As is mentioned in Section 2.2, for each treatment time three wells with bacterial suspension are considered, increasing thus the consistency of the results (i.e., CFU average is taken into account). On the top of that, as shown in Figure 3, two independent (carried out on different dates) series of experiments under identical conditions (see Section 2.2) are carried out, giving to the reader evidence for the standard deviation of the treatment efficiency. Clearly, the present plasma jet has a profound effect on the *A. baumannii* population which is drastically decreased mainly in the post-treatment time. The first measurement ("zero-time" point) is carried out 2 h after the treatment termination due to sample transfer and analysis waits. This first point (Figure 3) indicates colony diminution up to four orders of magnitude. The effect is more pronounced upon increasing bacteria exposure time to plasma. Cell death could thus be achieved either by increasing the plasma exposure dose or by allowing adequate time for the reactive species to initiate an intracellular signaling response in the cells [77].

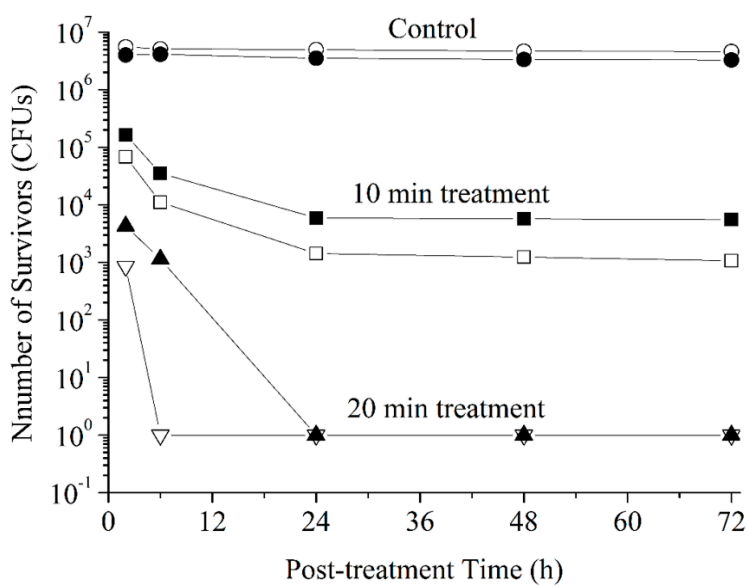

**Figure 3.** Post-treatment evolution of the *A. baumannii* formed colonies with the plasma treatment time as parameter. Each experimental point is derived by averaging the survivors from three different wells (see Section 2.2.), whereas open and closed symbols refer to two independent series of experiments realized under identical conditions, hence providing a picture for the standard deviation of the results.

Figure 4 depicts the evolution of the inhibition zone on MHA versus the treatment time, as it is observed 24 h after the treatment termination. An almost linear increase between the inhibition zone diameter and the treatment time is obtained (Figure 5). At the same time, in Figure 5 it is seen that as the plasma jet approaches the bacterial load, a wider inhibition zone is achieved. Although these experiments are not designed to provide quantitative results in terms of bacterium deactivation, an important conclusion may be extracted. The cross section of the observed plasma channel lies approximately up to 3 mm$^2$, whereas its visible spot at the point of contact with the specimen may be roughly 7–9 times wider in mm$^2$ (see e.g., Figure 1 in Ref. [31]). At the same time, the inhibition zone covers an area up to about 250 mm$^2$ (10 mm distance; 4 min treatment). Thus, the coverage area of the plasma-induced active species is much broader than the plasma visible spot and may be tuned by the working distance and the treatment time.

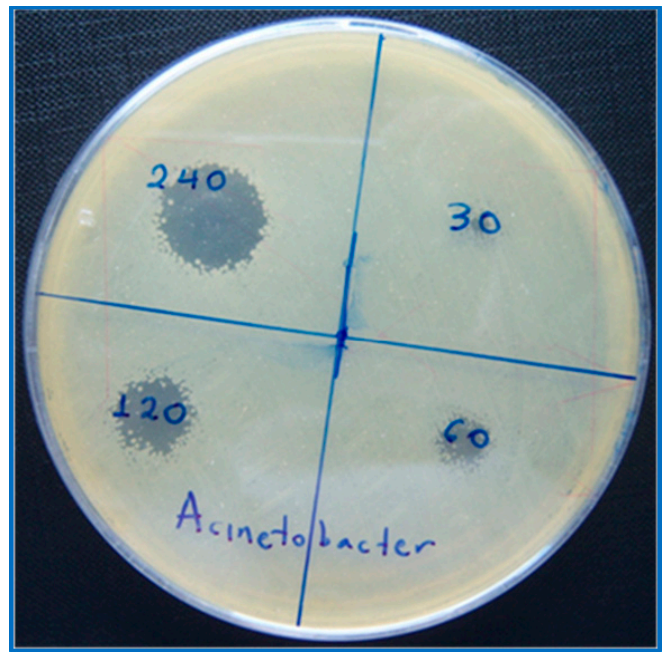

**Figure 4.** Demonstration of growth inhibition zones following treatment for various times (numbers on the image stand for seconds; the distance between the plasma-jet reactor orifice and the MHA surface is 15 mm).

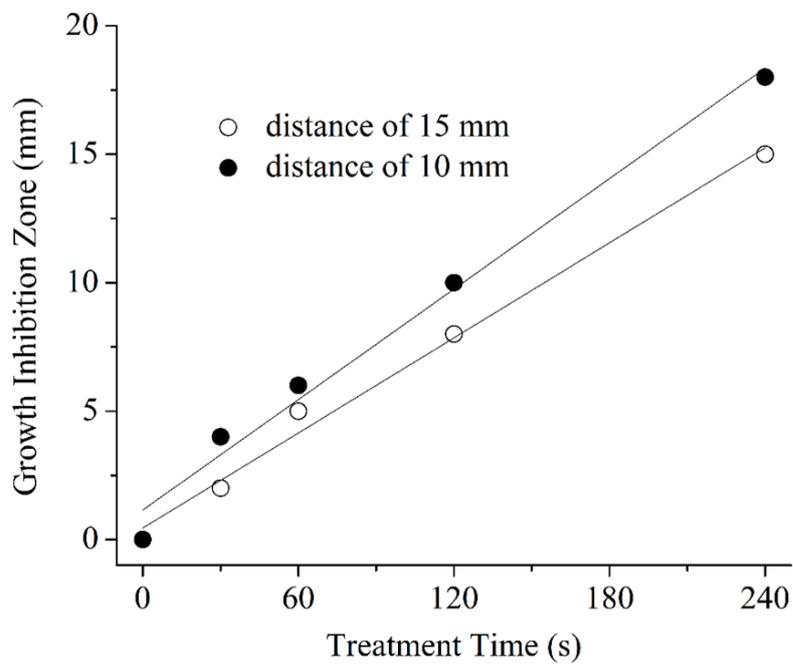

**Figure 5.** Diameter of the growth inhibition zone as a function of the plasma treatment time, for two different distances between the reactor orifice and the MHA surface. Duplicated experiments are considered, and the standard deviation is found to be around 10%. The lines refer to linear fitting of the experimental data (open and solid symbols).

Untreated and plasma-treated bacteria as probed by SEM are shown in Figure 6. Treated bacteria are significantly fewer and are completely deformed. More specifically, they show ruptured and shrunk morphology. It cannot be stated whether this morphological degradation is the cause or the result of the bacteria death. In any case the plasma jet leads to leakage of intracellular ingredients leaving empty cell walls and membranes. No correlative evidence has been observed [51] suggesting that plasma induced widespread genomic damage or oxidative protein modification prior to the onset of membrane damage.

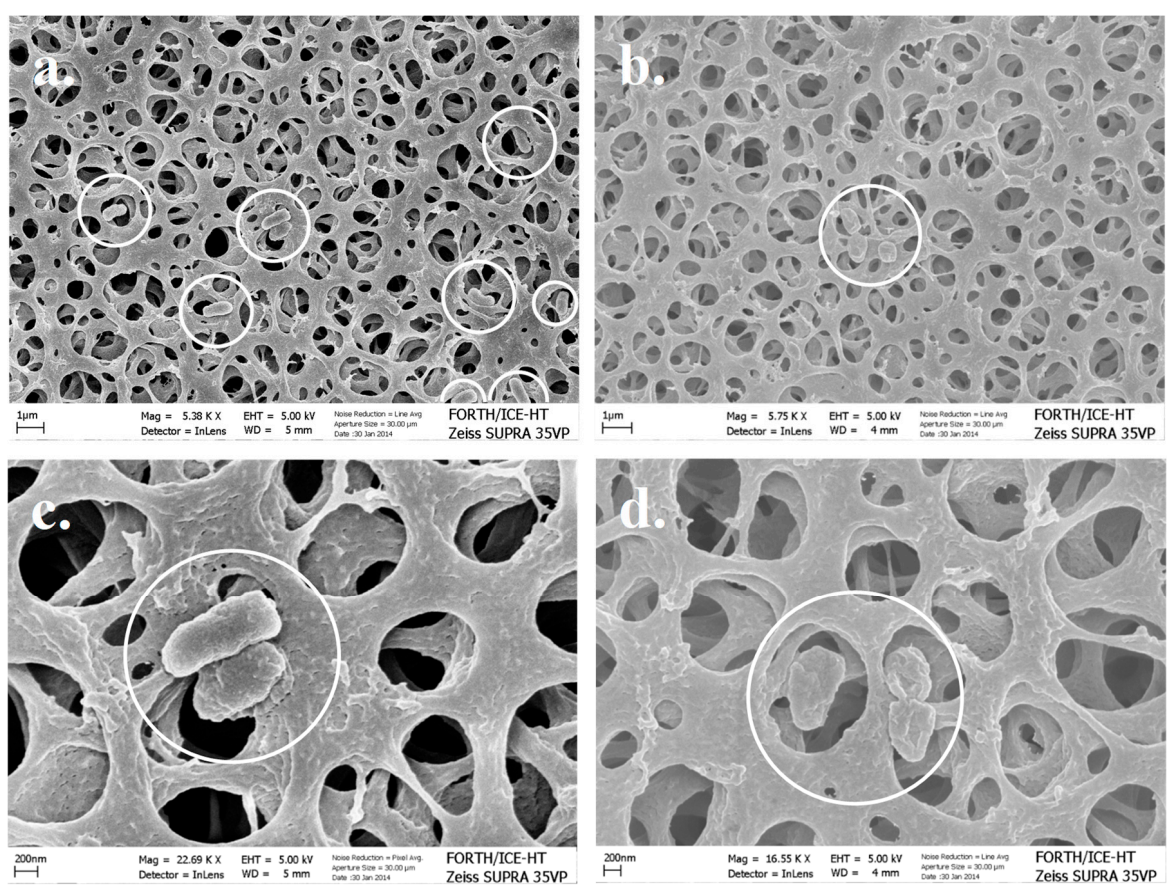

**Figure 6.** SEM micro-images of *A. baumannii* bacteria (**a**) untreated-control and (**b**). plasma-treated for 20 min. Images were taken about 24 h after the treatment of the samples (refer to Figure 3). Both control and treated samples were stored at 4 °C. (**c**,**d**) Corresponding zoomed-in micro-images (control and treated, respectively).

According to the wide scan optical emission spectrum of the present plasma jet under the same operating conditions (as it is analyzed elsewhere [29,30]), the aqueous solution with the bacteria is subjected to a flow of (re)active species including: $N_2(C^3\Pi_u–B^3\Pi_g)$, $N_2^+(B^2\Sigma^+_u–X^2\Sigma^+_u)$, $OH(A^2\Sigma^+–X^2\Pi)$, $He(3^1P–2^1S; 3^3D–2^3P; 3^1D–2^1P; 3^3S–2^3P)$, and $NO_\beta(B^2\Pi–X^2\Pi)$ of much weaker intensity. At the same time, the gas temperature of the free-running plasma-jet is evaluated based on the rotational distributions of typical probe molecules (Figure 7). The measured temperature lies between 54 and 65 °C. When higher resolution spectra are recorded [30], a temperature around 36 °C is found. Thus, the role of heating in *A. baumannii* inactivation with the present setup should not be considered the principal role, in agreement with other reports [64,65].

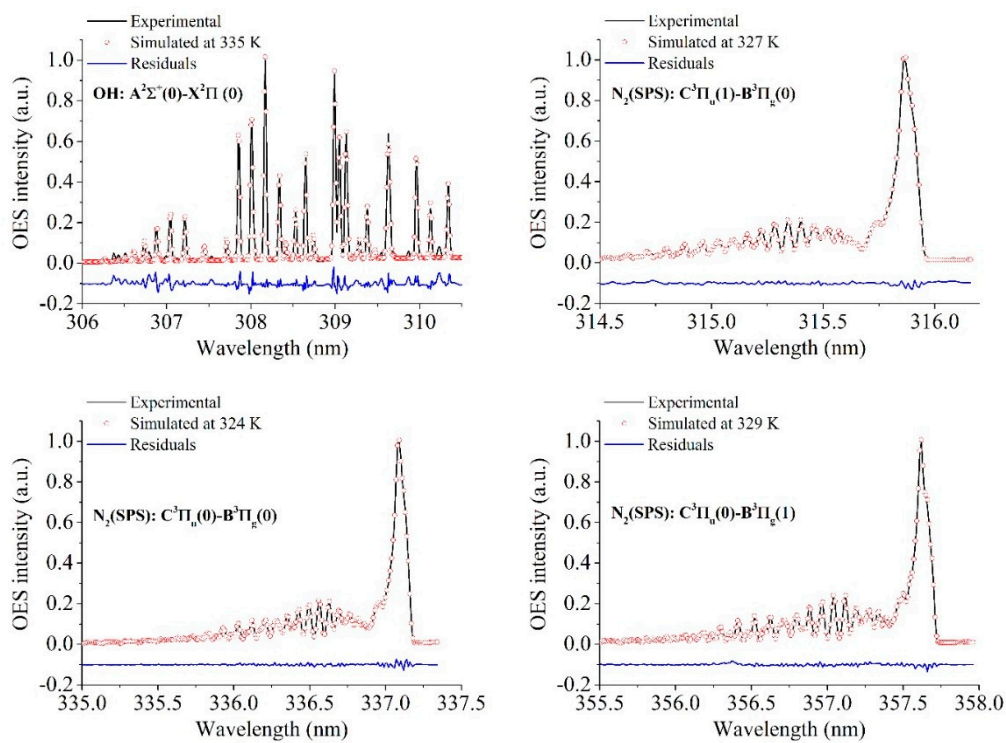

**Figure 7.** Theoretical rotational distributions fitted on experimentally determined ones for the estimation of the gas temperature (the residuals of the fitting procedure are also plotted). The probe molecules and the corresponding transitions are given in the insets.

It is widely accepted that one single agent cannot be assigned to the plasma sterilization of bacteria. It is suggested that several germicidal species that include RONS [33,64,65], charged particles [48,64], UV radiation [47], ozone molecules [37], metastables [78] and electric field [16] may have a potential role. Together, they lead, possibly in combination, to inactivation of bacteria [63]. Furthermore, different sterilizing mechanisms proceed at different timescales. This assumption is supported by various published survival curves [79]. These curves describe the dependence of the number of surviving bacteria on the treatment time. Many of these curves exhibit a double- or multi-slope shape, like in our case (Figure 3). Such a behavior is indicative of the relevance of several distinct sterilization mechanisms [63]. Bactericidical action of plasmas may be associated with (i) wall permeabilization by electromechanical and chemical processes (including electrostatic disruption of Gram-negative bacteria [80] like *A. baumannii*), (ii) cell penetration of reactive oxygen and nitrogen species and (iii) chemical reactions inside the cell that can lead to bacterial DNA destruction [81].

On the other hand, the above mechanisms may be direct on the bacteria or indirect by altering the chemical properties of the aqueous solution where the bacteria are suspended. For instance, ozone decomposition in water is complex and many transient oxidizing species (OH, $HO_2$, $O^-$, $O_3^-$, O and singlet $^1O_2$) can be formed [82]. We have recently shown the formation of reactive nitrogen species including peroxynitrite in a physiological buffer exposed to He APPJ [28]. Girard et al. [27], using a micro-plasma jet produced in helium, demonstrated that the concentration of $H_2O_2$, $NO_2^-$ and $NO_3^-$ can fully account for the majority of RONS produced in plasma-activated buffer, leading finally to tumor cell death. Ercan et al. [69] used floating-electrode DBD and they stated that nonthermal plasma-treated fluids retain their antimicrobial effects for longer periods (three months by delay time and two years by solution aging) than was earlier thought. This fluid-mediated plasma-based treatment is faster than reported previously (<15 min holding time, defined as the time that plasma-treated liquid comes in contact with the bacterial suspension) and inactivates a wide range of multidrug-resistant bacteria and fungal pathogens, in both their planktonic and biofilm forms. The antimicrobial effect is due not only to a change in pH and in the $H_2O_2$ or nitric acid but also to the likely generation of additional species or products that are responsible for the powerful biocidal effect [69].

## 4. Conclusions

Under the conditions of the present work, *A. baumannii* may efficiently be deactivated by atmospheric-pressure cold plasma produced in the form of "guided streamers" which propagate within helium channeling into atmospheric air. The reported results demonstrated that the application of this plasma-based method could contribute to restriction of hospital-associated infections and antibiotic use. Specifically: (i) inhibition zone experiments in agar mirrored the potential action of the present system against *A. baumannii* in its planktonic or biofilm mode of growth (e.g., catheter surface sterilization), whereas (ii) colony count results from the treatment of *A. baumannii* suspensions implied the possibility of activating aqueous solutions and their latter use for curing infected areas. In the first 24 h after a 20 min treatment, the formation of *A. baumannii* colonies was interrupted completely and the bacteria were morphologically deformed. *A. baumannii* eradication did not involve high temperature phases and should principally be triggered by the plasma-induced reactive species. Finally, the direct application (physical contact) of the tested plasma jet on patients cannot be justified by the presently available data, in terms of safety.

**Author Contributions:** For the present work the authors' contribution may be distinguished as follows. P.S.: conceptualization; methodology; validation; investigation; resources; writing; supervision; project administration. A.S.: methodology; validation; investigation; resources; writing. P.G.K.: methodology; investigation; resources; writing. K.G.: formal analysis; investigation; writing; visualization. E.D.A.: conceptualization; methodology; resources; supervision.

**Funding:** This research received no external funding.

**Conflicts of Interest:** The authors declare no conflict of interest.

*Plasma* **2019**, *2* 89